\documentclass[runningheads]{llncs}
\usepackage[T1]{fontenc}
\usepackage{graphicx}
\usepackage{amsmath}
\begin{document}
\title{Efficient Fine-Tuning of DINOv3 Pretrained on Natural Images for Atypical Mitotic Figure Classification}
\titlerunning{DINOv3 PEFT for MIDOG 2025 ANMF Classification}
%
\author{Guillaume Balezo\inst{1, 2, 3}\orcidID{0000-0002-7750-8852} \and
Raphaël Bourgade\inst{1, 4, 5, 6}\orcidID{0009-0000-6945-7170} \and
Hana Feki\inst{1}\orcidID{0009-0005-4519-3103} \and
Lily Monnier\inst{1, 4, 5}\orcidID{0009-0009-4346-8301} \and
Matthieu Blons\inst{1, 4, 5}\orcidID{0000-0002-8609-3118} \and
Alice Blondel\inst{1, 4, 5}\orcidID{0000-0002-0342-4664} \and
Etienne Decencière\inst{2}\orcidID{0000-0002-1349-8042} \and
Albert Pla Planas\inst{3}\orcidID{0000-0001-7527-2500} \and
Thomas Walter\inst{1, 4, 5}\orcidID{0000-0001-7419-7879}
}
\authorrunning{G. Balezo et al.}
%
\institute{
Mines Paris, PSL University, Center for Computational Biology, 75006 Paris, France \and
Mines Paris, PSL University, Center for Mathematical Morphology, 77300 Fontainebleau, France \and
Sanofi, Paris, France \and
Institut Curie, PSL University, Paris, France \and
INSERM, U1331 Computational Oncology, Paris, France \and
Department of Pathology, University Hospital of Nantes, Nantes, France
}
\maketitle              
\begin{abstract}
Atypical mitotic figures (AMFs) indicate abnormal cell division associated with poor prognosis. Their detection remains difficult due to low prevalence, subtle morphology, and inter-observer variability. The MItosis DOmain Generalization (MIDOG) 2025 challenge introduces a benchmark for AMF classification across multiple domains. In this work, we fine-tuned the recently published DINOv3-H+ vision transformer, pretrained on natural images, using low-rank adaptation (LoRA), training only 1.3M parameters. We combine this with extensive augmentation and a domain-weighted Focal Loss to better handle the strong domain heterogeneity in the dataset. Despite the large shift between natural images and histopathology, our fine-tuned DINOv3 transfers effectively, reaching first place on the final test set. These results highlight the advantages of DINOv3 pretraining and underline the efficiency and robustness of our fine-tuning strategy, yielding state-of-the-art results for the atypical mitosis classification challenge in MIDOG 2025. Our code is publicly available on GitHub\footnote{\url{https://github.com/Sanofi-Public/EFTD-midog-amf}}.

\end{abstract}
\section{Introduction}

Mitotic activity is a central indicator of tumor proliferation and prognosis. Beyond simple counts, the distinction between normal mitotic figures (NMFs) and atypical mitotic figures (AMFs) is of particular interest, as AMFs reflect abnormal cell division processes and correlate with poor clinical outcomes. However, their identification is challenging due to low prevalence, subtle morphological differences, and low inter-rater agreement even among trained pathologists. Automated image analysis methods therefore have the potential to improve reproducibility and reduce observer bias in this task.

The MItosis DOmain Generalization Challenge 2025 (MIDOG 2025) \cite{midog2025} extends the scope of previous editions (MIDOG 2021 \cite{midog2021} and MIDOG 2022 \cite{midog2022}) with the goal of advancing robust AI-assisted cancer diagnosis. Task 2 introduces a dedicated benchmark for AMF classification using H\&E (Hematoxylin and Eosin) images. Participants were asked to classify cropped cell patches (128×128 pixels) into NMF or AMF across multiple tumor types, species, scanners, and laboratories. The dataset contains more than 12,000 annotated mitotic figures, with AMFs accounting for only $\sim$20\% of cases. The evaluation metric of the challenge is the balanced accuracy to mitigate this strong class imbalance. Similar to the earlier MIDOG challenges, this benchmark addresses the crucial problem of robustness and generalization across domains, now extended to the clinically relevant task of atypical mitosis classification.

In this work, we tackle Task 2 by applying low-rank adaptation (LoRA) \cite{lora} to fine-tune DINOv3-H+-LVD-1689M, a vision transformer (ViT) pretrained on natural images using the state-of-the-art DINOv3 self-supervised (SSL) method \cite{dinov3}. Recent progress suggests that such generic foundation models, though developed outside the biomedical domain, can be efficiently adapted to specialized medical imaging tasks. To enhance robustness across diverse domains and compensate for the limited number of atypical figures, we combine this strategy with extensive data augmentation and a Domain Weighted Focal Loss. Our approach aims to test and leverage the representational power of this generic DINOv3 SSL pretraining, while ensuring efficient adaptation to the histology-specific and heterogeneous challenge dataset. In parallel, our team also explored a ConvNeXt-based solution that we present in \cite{hana-convnext}. It achieved good performance, though slightly below DINOv3 on the preliminary test set. We therefore retained DINOv3 as our final solution, which achieved the first place on the final leaderboard.

\section{Material and Methods}

\subsection{Dataset}

\begin{figure}
\centering
\includegraphics[width=\linewidth]{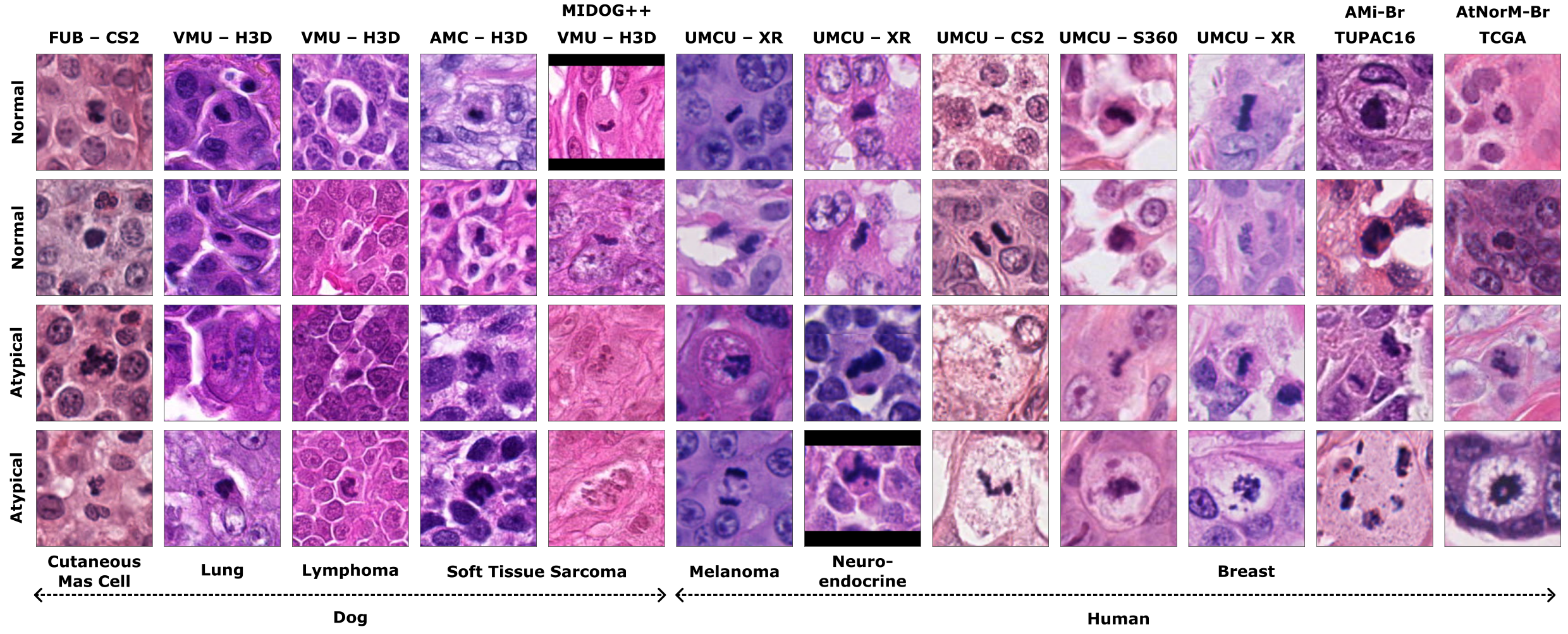}
\caption{\textbf{Training samples across domains:} Examples of normal and atypical mitotic figures across diverse domains, including variations in species (dog, human), organs, centers, and scanners (CS2, H3D, XR, S360). Each column corresponds to one domain, highlighting the strong appearance shifts present in the training data.}
\label{fig:dataset}
\end{figure}

We describe the datasets used for training and evaluation, including the official MIDOG 2025 data and three external sources added to improve domain diversity.

The \textbf{MIDOG 2025} \cite{midog2025} atypical mitosis dataset is derived from 454 histopathology images spanning nine domains defined by different tumor types, species, scanners, and laboratories. Each mitotic figure was subtyped as normal or atypical by three expert pathologists in a blinded majority-vote setting.

In addition to the official MIDOG 2025 atypical dataset, we incorporated three external resources. The \textbf{AMi-Br} \cite{amibr} dataset provides mitotic figures from MIDOG 2021 \cite{midog2021} and TUPAC16 \cite{tupac16}. To avoid overlap with the main dataset, we used only the TUPAC16 subset. The \textbf{AtNorM-Br} \cite{atnormbr} dataset contains mitotic figures from the TCGA \cite{tcga} breast cancer cohort, annotated by an expert pathologist. Finally, the \textbf{OMG-Octo} dataset \cite{omg-octo} was created by screening large histopathology data with a mitosis detection model pretrained on AMi-Br and MIDOG25, followed by expert review of candidate mitoses. Figure~\ref{fig:dataset} illustrates representative examples from these domains, highlighting the large visual variability across datasets.

After removing duplicate images across datasets, we constructed the final training set used for model development. A detailed breakdown by dataset and class is provided in Table~\ref{tab:dataset_summary}. All datasets were provided as 128×128 pixel crops centered on the mitotic figure, except for OMG-Octo, which was originally 64×64 pixels and resized to 128×128 for training, corresponding to a resolution of $0.25\,\mu\text{m/pixel}$.

\begin{table}[h]
\centering
\caption{Mitotic figure counts in the final training set, by dataset and class.}
\begin{tabular}{l@{\hspace{10pt}}c@{\hspace{10pt}}c@{\hspace{10pt}}c}
\hline
\textbf{Dataset} & \textbf{Normal} & \textbf{Atypical} & \textbf{Total} \\
\hline
MIDOG 2025           & 10,191 & 1,748 & 11,939 \\
AMi-Br (TUPAC16)     & 1,571  & 428   & 1,999  \\
AtNorM-Br            & 587    & 124   & 711    \\
OMG-Octo             & 378    & 1,374 & 1,752  \\
\textbf{Total}       & \textbf{12,724} & \textbf{3,674} & \textbf{16,398} \\
\hline
\end{tabular}
\label{tab:dataset_summary}
\end{table}

The preliminary test set provided for Task 2 consisted of mitotic figure crops from four tumor types not included in the final test data. It was made available on the challenge platform two weeks prior to submission for debugging purposes. The final test set consists of patches from 120 cases covering 12 distinct tumor types from both human and veterinary pathology, with 10 cases per tumor type. This set spans multiple laboratories and scanning systems and was used for the official evaluation. Performance was assessed using balanced accuracy, computed over all patches of the test set.

\subsection{Methods}

\begin{figure}
\centering
\includegraphics[width=\linewidth]{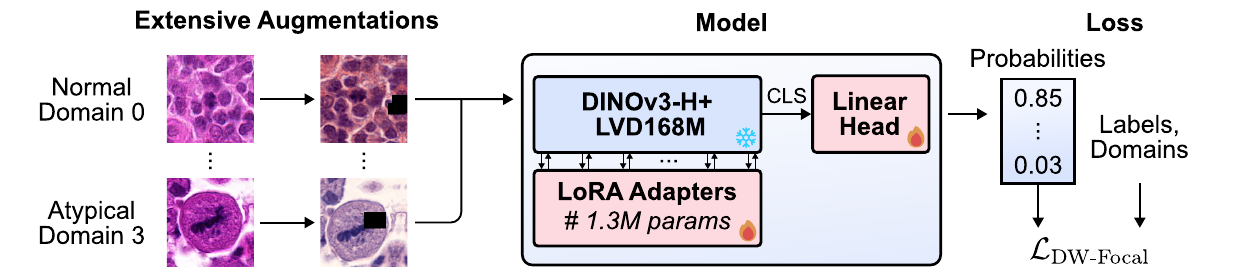}
\caption{\textbf{Overview of our method during training:} Input images are augmented (multi-Macenko, small translations, shear, coarse dropout, rotations, etc.) and normalized with ImageNet statistics. The classifier is a DINOv3-H+ pretrained on the LVD-1689M natural image dataset, fine-tuned with LoRA (rank 8, $\alpha=16$, $\sim$1.3M trainable parameters) and followed by a linear head on the class token with sigmoid activation to output probabilities. Optimization is performed with a Domain-Weighted Focal Loss, which combines Focal Loss for class imbalance with domain reweighting to address dataset heterogeneity.}
\label{fig:main_workflow}
\end{figure}

In this section, we detail the main components of our workflow (Figure~\ref{fig:main_workflow}): network training setup, the proposed Domain-Weighted Focal Loss, data augmentation strategy, and test-time augmentation.

\subsubsection{Network Training}
We trained our model on 128$\times$128 pixel image crops, matching the original patch size defined in the challenge. Our model is a DINOv3-H+ ViT pretrained on the LVD-1689M natural image dataset. We fine-tuned it for Task 2 using LoRA with $rank=8$, $\alpha_{\text{LoRA}}=16.0$, and dropout 0.05, applied only to the query and value projections in the attention layers. This resulted in just 1.3M trainable parameters. A linear classification head with 0.2 dropout and a sigmoid activation was added to produce probabilities from the class token. Training was run with a batch size of 16 and mixed precision (FP16), accelerating training and reducing memory usage without impacting predictive performance. We used the AdamW optimizer with a learning rate of $1 \times 10^{-4}$, a weight decay of 0.1, and $\epsilon=1 \times 10^{-7}$. A cosine learning rate schedule with linear warmup was applied over the first 10\% of training starting from $8.47 \times 10^{-7}$ to the base learning rate. Gradient norms were clipped at 1.0 for stability. Inputs were normalized with ImageNet statistics, consistent with DINOv3 pretraining. The final submitted model was trained for 60 epochs using the full combined dataset: AMi-Br (TUPAC16), MIDOG25, AtNorM-Br, and OMG-Octo.

\subsubsection{Domain-Weighted Focal Loss}
To address class imbalance ($\sim$20\% atypical) and domain heterogeneity, we used an extended version of the Focal Loss \cite{focal_loss} with $\alpha_{y_i=1}=0.25$, $\alpha_{y_i=0}=0.75$, and $\gamma=2$. This extension, which we named Domain Weighted (DW) Focal Loss, incorporates domain reweighting: domain weights are first computed as the inverse square root of domain size. Then to avoid instability due to large values, we cap the ratio between the largest and smallest weights at 3. Finally, we normalize the adjusted weights to sum up to 1.

\begin{equation}
\mathcal{L}_{\text{DW-Focal}} = - \frac{1}{N} \sum_{i=1}^N 
w_{d(i)}\, \alpha_{y_i}\, (1 - p_{i,y_i})^{\gamma} \log p_{i,y_i},
\end{equation}

where $N$ is the number of samples, $p_{i,y_i}$ is the predicted probability for the ground-truth class $y_i$ ($y_i=1$ for NMF and $y_i=0$ for AMF), and $w_{d(i)}$ is the weight of sample $i$’s domain $d(i)$, computed as:

\begin{equation}
u_d = n_d^{-1/2},\;
v_d = \min(u_d,\, 3 \cdot \min_j u_j),\;
w_d = \frac{v_d}{\sum_j v_j},
\end{equation}

where $n_d$ is the number of samples in domain $d$.

Indeed, the Focal Loss reduces the relative contribution of well-classified (easy) samples and forces the model to focus more on uncertain or difficult ones. In addition, the more frequent class is down-weighted by the choice of $\alpha$. This gives more weight to the minority class and influences the balance between recall and precision. The third aspect of the domain-weighted Focal Loss is to also assign more weight to underrepresented domains.

\subsubsection{Data Augmentation}
We also applied extensive data augmentations, including color jitter, JPEG compression, stain augmentation (multi-Macenko \cite{multi_macenko,macenko} with random stain domain references), defocus blur, affine transforms, D4 symmetry, coarse dropout (up to two random boxes of 10–32 pixels per side), and a custom black-border augmentation to mimic zero-padded regions in the training data.

We converted the \textbf{multi-Macenko} normalizer \cite{multi_macenko} into augmentation by first extracting stain matrices from 10 randomly selected images per domain, using both the training set and additional cases from MITOS CMC \cite{mitos_cmc}, MITOS CCMCT \cite{mitos_ccmct}, and TCGA COAD/BLCA \cite{tcga}. This was done to better address domain shift, especially from unseen domains in the final test set, by simulating a wider range of staining variations. During training, we sampled a domain and a random subset of these references, averaged their stain matrices as proposed in \cite{multi_macenko}, and used this as the target for Macenko normalization. To mimic staining variability, we further perturbed the normalized images in stain space with additive and multiplicative uniform noise ($\sigma=0.2$).

\subsubsection{Test Time Augmentation}
At inference, we employed test-time augmentation by averaging logits over four rotated views (0°, 90°, 180°, and 270°) to improve robustness.

\section{Evaluation and Results}

\begin{figure}
\centering
\includegraphics[width=\linewidth]{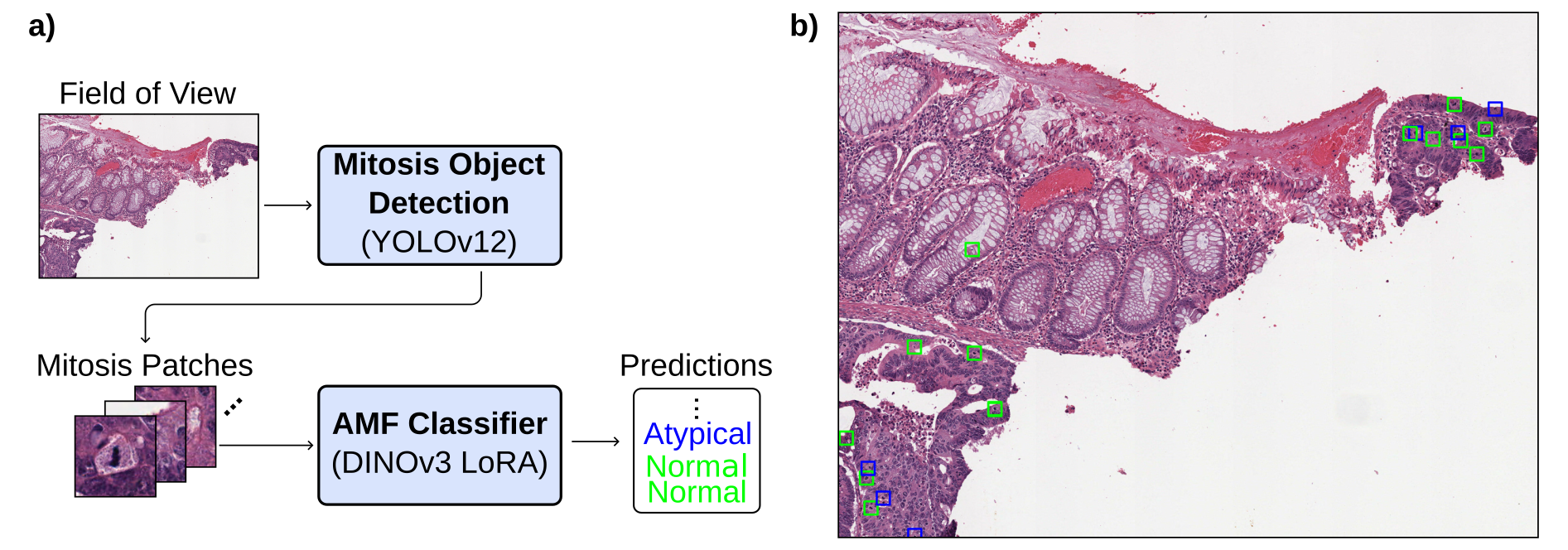}
\caption{\textbf{Possible whole-slide inference workflow for mitosis analysis.}
\textbf{a)} Overview of a two-stage pipeline: a pre-existing mitosis detector is first applied to a field of view (FoV) to localize mitotic figures. Detections are cropped into 128×128 patches, which are then classified as normal or atypical using our DINOv3-LoRA classifier. Predictions can be mapped back to the FoV to reconstruct labeled bounding boxes.
\textbf{b)} Example inference on a unseen colorectal FoV from TCGA, showing normal (green) and atypical (blue) mitosis predictions.}
\label{fig:wsi_pred}
\end{figure}

\subsection{Overall Performance}
We evaluated our DINOv3-H+-based method, using 4-fold cross-validation on the training data, holding out the AMi-Br TUPAC16 subset as an external test set. In each fold, we measured balanced accuracy (BA) on the held-out validation split and on AMi-Br TUPAC16, and reported the mean $\pm$ standard deviation across folds. We also report the performances of the submitted DINOv3-H+ model (trained on the full training set) on both the preliminary and the final test sets, which are non-overlapping (Table~\ref{tab:main_perfs}). For comparison, we included DINOv3-ConvNeXt-Tiny, a distilled variant pretrained on LVD-1689M and used here as a strong lightweight baseline, as suggested by our second approach~\cite{hana-convnext}. Unlike our LoRA-based approach, DINOv3-ConvNeXt-Tiny was fully fine-tuned with a dropout of 0.2, leading to about 21× more trainable parameters despite having fewer overall parameters. Overall, DINOv3-H+ with LoRA consistently outperformed the ConvNeXt baseline across all evaluations. The performance gap between validation and the external AMi-Br TUPAC test set was also smaller, likely because full fine-tuning of ConvNeXt-Tiny led to stronger overfitting to the training domain.

\begin{table}[h]
\centering
\caption{Performance on 4-fold cross-validation (mean $\pm$ std), the external AMi-Br (TUPAC) test set, and the preliminary and final test sets$^*$. 
$^*$Evaluated with a model trained on the full training data (outside CV).}
\begin{tabular}{l@{\hspace{10pt}}c@{\hspace{10pt}}c}
\hline
\textbf{Split} & \textbf{DINOv3-H+} & 
\begin{tabular}{c}
\textbf{DINOv3-} \\
\textbf{ConvNeXt-Tiny}
\end{tabular} \\
\hline
Cross-Validation & \textbf{0.9485 $\pm$ 0.0038} & 0.9352 $\pm$ 0.0032 \\
AMi-Br (TUPAC16)     & \textbf{0.8475 $\pm$ 0.0060} & 0.8085 $\pm$ 0.0034 \\
Preliminary Test$^*$ & \textbf{0.9045} & - \\
Final Test$^*$       & \textbf{0.9079} & - \\
\hline
\end{tabular}
\label{tab:main_perfs}
\end{table}

\subsection{Effect of Domain Reweighting Loss}
Replacing Focal Loss with Domain-Weighted Focal Loss improved mean domain-balanced accuracy in cross-validation (0.9284 → 0.9341) and yielded a notable gain on the preliminary test set (0.8870 → 0.9045), demonstrating increased robustness across domains. On the AMi-Br test set, however, performance remained almost unchanged (0.8467 → 0.8475). It is likely due to a stronger and unseen domain shift between this specific test and training domains, which could not be addressed by reweighting since it only balances training domains (Table~\ref{tab:loss_perfs}).

\begin{table}[h]
\centering
\caption{Balanced accuracy (BA) with Focal vs. Domain-Weighted Focal (DW-Focal) losses. 
CV reports mean BA across domains ($\pm$ std). 
AMi-Br and Preliminary Test report overall BA. 
$^*$Model trained on full available data.}
\begin{tabular}{l@{\hspace{10pt}}c@{\hspace{10pt}}c}
\hline
\textbf{Split} & \textbf{Focal} & \textbf{DW-Focal} \\
\hline
\multicolumn{3}{c}{\textit{Mean BA across domains}} \\
\hline
Cross-Validation & 0.9284 $\pm$ 0.0031 & \textbf{0.9341 $\pm$ 0.0057} \\
\hline
\multicolumn{3}{c}{\textit{Overall BA}} \\
\hline
AMi-Br (TUPAC16) & 0.8467 $\pm$ 0.0060 & \textbf{0.8475 $\pm$ 0.0060} \\
Preliminary Test$^*$ & 0.8870 & \textbf{0.9045} \\
\hline
\end{tabular}
\label{tab:loss_perfs}
\end{table}

\subsection{SSL Finetuning on Mitosis Images}
We also explored continuing the self-supervised training of the DINOv3-L-LVD-1689M model on mitosis-like images. Using a mitosis detector trained on the Task 1, we collected candidate patches from TCGA BLCA/COAD, MITOS CMC/CCMCT, and the Task 2 training data. This resulted in a collection of approximately 260k image crops (128×128). Using the official DINOv3 pipeline, we performed self-supervised fine-tuning of LoRA parameters, starting from LVD-1689M-pretrained weights. We then performed linear probing on the full Task 2 dataset to evaluate the learned representations, following standard practice in SSL. Despite being constrained to 4 GPUs and a reduced global batch size of 64, the SSL fine-tuning yielded a $\sim$10\% improvement in balanced accuracy, from 53.44\% with original DINOv3-L-LVD-1689M to 63.11\%. As expected, this suggests that large-scale SSL on mitosis images could be beneficial. However, due to time and budget constraints, we did not fully investigate DINOv3 pretraining on histology images, leaving it as a promising direction for future work.

\begin{table}[h]
\centering
\caption{Balanced accuracy from linear probing on Task 2, before and after SSL fine-tuning on mitosis-like images.}
\begin{tabular}{l@{\hspace{10pt}}c@{\hspace{10pt}}c}
\hline
\textbf{Model} & \textbf{LVD-1689M} & \textbf{SSL Finetuned (LoRA)} \\
\hline
DINOv3-L & 0.5344 & \textbf{0.6311} \\
\hline
\end{tabular}
\label{tab:ssl_perfs}
\end{table}

\subsection{Out-of-Domain TCGA Analysis}

\begin{figure}
\centering
\includegraphics[width=\linewidth]{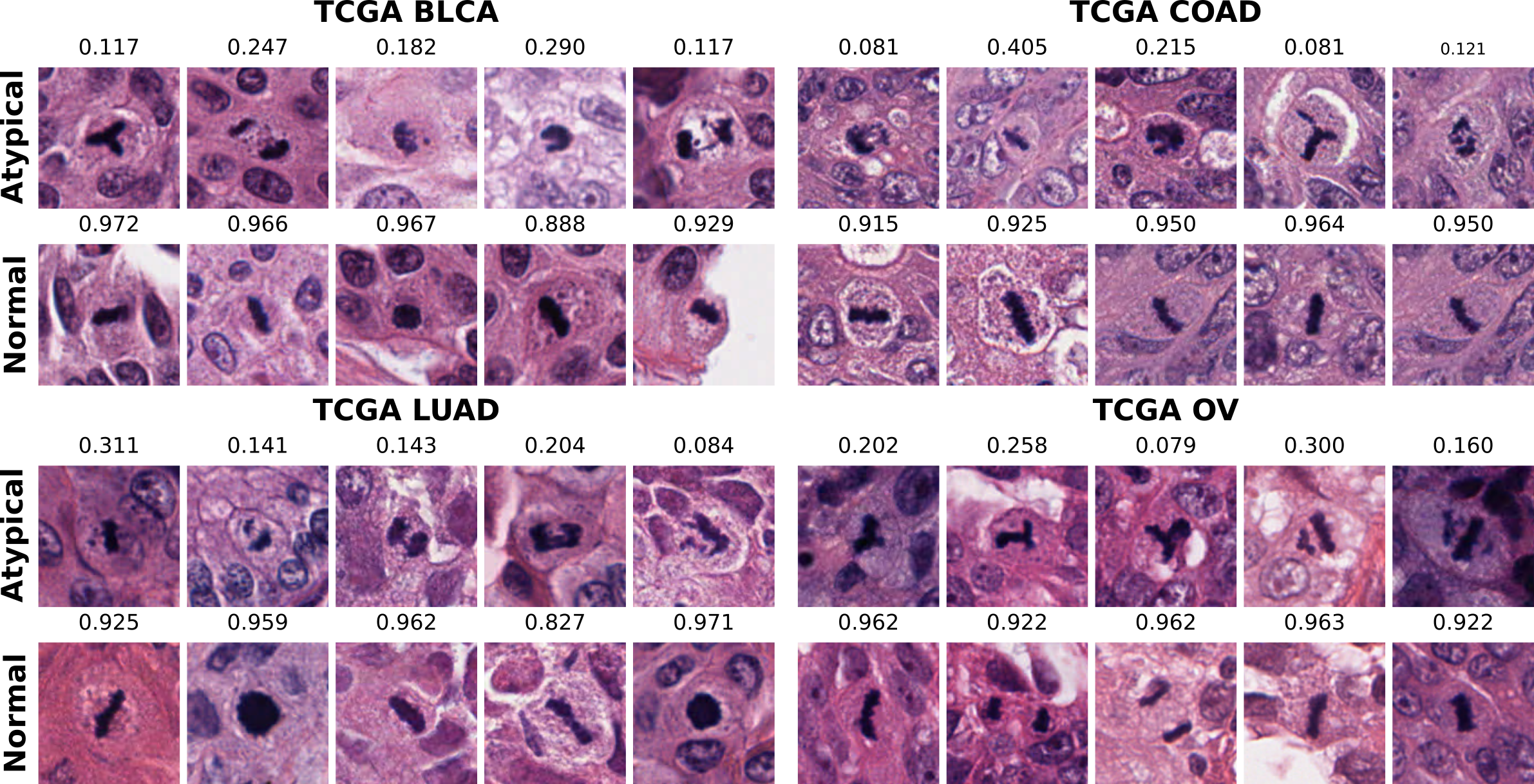}
\caption{\textbf{Representative predictions on external TCGA cohorts:} COAD (colorectal), BLCA (bladder), OV (ovarian), LUAD (lung). Images were reviewed and annotated by an expert pathologist, highlighting correct normal and atypical mitosis classifications along with their predicted probabilities. A high probability indicates a normal mitosis and low probability indicates an atypical mitosis.}
\label{fig:tcga_vis}
\end{figure}

Because the preliminary and final test sets remain private, and our model was trained on the fully available dataset, we conducted an additional qualitative (visual-only) out-of-domain evaluation. This was done on TCGA, across four organs not included in the training set: liver, ovarian, colorectal, and bladder. Following the workflow depicted in Figure~\ref{fig:wsi_pred}, we first applied our YOLOv12 mitosis detector \cite{task1} (developed for Task 1) to extract candidate mitotic figures, and then classified each crop using our DINOv3 ANMF classifier.

An expert pathologist reviewed a diverse set of predictions and compiled representative correct and incorrect cases, shown in Figure~\ref{fig:tcga_vis}. Consistent with the strong performance observed on the final test set, the model
exhibits good qualitative generalization to unseen organs and staining domains.

\section{Discussion}

In this work, we demonstrated that DINOv3-H+ with LoRA fine-tuning constitutes a leading approach for atypical mitosis classification in MIDOG 2025, training only 1.3M parameters. Despite the domain gap, the LVD-1689M-pretrained DINOv3 transferred well to histopathology, thanks to the efficacy of our fine-tuning strategy, achieving first place on the final test set. This contributes in proving that generalist foundation models, even when trained on natural images, can capture meaningful patterns useful for biomedical imaging tasks and support specialized applications such as atypical mitosis classification.

DINOv3-H+ remains a large model, which can be costly at inference, especially since mitosis classification often follows object detection with hundreds of thousands of candidate patches. Our parallel ConvNeXt approach underlined that smaller networks can reach near state-of-the-art performance while requiring fewer overall parameters. Future directions include leveraging knowledge distillation to develop more efficient models. Stronger augmentations, such as optical flow or grid distortion, and increasing LoRA capacity (higher rank or more layers) could also improve adaptation to histology images.

Another promising direction is to build stronger integration with the Task 1 of the challenge. This could involve adding hard negative mitoses to the training set, exploring DINOv3 as a backbone for mitosis detection given its strong performance on dense downstream tasks, or applying our classification approach on top of a mitosis detector (Task 1) to further improve performance. Future work may also explore DINOv3 SSL pretraining on histopathology images, especially mitosis-like patches. Such extensions could further improve generalization across domains and strengthen the role of foundation models for mitosis subtyping.

\begin{credits}
\subsubsection{\ackname} The authors gratefully acknowledge the MIDOG organizers and the Grand Challenge platform for hosting the challenge and providing access to such valuable datasets.

This work was supported by the French government under the management of Agence Nationale de la Recherche as part of the “Investissements d’avenir” program, reference ANR-19-P3IA-0001 (PRAIRIE 3IA Institute) and ANR-23-IACL-0008 (PRAIRIE-PSAI). Furthermore, this work was supported by a government grant managed by the Agence Nationale de la Recherche under the France 2030 program, with the reference numbers ANR-24-EXCI-0001, ANR-24-EXCI-0002, ANR-24-EXCI-0003, ANR-24-EXCI-0004, ANR-24-EXCI-0005.

This work was also supported by Sanofi through a CIFRE PhD fellowship in collaboration with Mines Paris – PSL University, with the support of the Association Nationale de la Recherche et de la Technologie (ANRT).

\subsubsection{CRediT authorship contribution statement}

\textbf{Guillaume Balezo}: Conceptualization; Methodology; Software; Investigation; Data curation; Validation; Visualization; Project administration; Writing – original draft; Writing – review \& editing.

\textbf{Raphaël Bourgade}: Conceptualization; Methodology; Data curation; Project administration; Validation; Visualization; Writing – review \& editing.

\textbf{Hana Feki}: Conceptualization; Methodology; Software; Investigation; Data curation; Validation; Writing – review \& editing.

\textbf{Lily Monnier}: Conceptualization; Data curation; Writing – review \& editing.

\textbf{Matthieu Blons}: Conceptualization.

\textbf{Alice Blondel}: Conceptualization; Project administration; Writing – review \& editing.

\textbf{Etienne Decencière}: Resources; Supervision; Project administration; Writing – review \& editing.

\textbf{Albert Pla Planas}: Resources; Supervision; Project administration; Writing – review \& editing.

\textbf{Thomas Walter}: Conceptualization; Resources; Supervision; Project administration; Writing – review \& editing.

\subsubsection{\discintname}
Authors 1 and 8 are Sanofi employees and may hold shares and/or stock options in the company.
\end{credits}
%
%
%
%
\bibliographystyle{splncs04}
\bibliography{literature}
\end{document}